\documentclass{ws-procs9x6}

\newcommand{\be} {\begin{eqnarray*}}
\newcommand{\ee} {\end{eqnarray*}}

\newcommand{\bcen}{\begin{center}}
\newcommand{\ecen}{\end{center}}
\newcommand{\beq}{\begin{equation}}
\newcommand{\eeq}{\end{equation}}
\newcommand{\bea}{\begin{eqnarray}}
\newcommand{\eea}{\end{eqnarray}}
\newcommand{\ba}{\begin{array}}
\newcommand{\ea}{\end{array}}
\newcommand{\bann}{\begin{eqnarray*}}
\newcommand{\eann}{\end{eqnarray*}}

\begin{document}

\title{ Energy release due to antineutrino untrapping from hot quark 
stars
\footnote{\uppercase{C}onference \uppercase{P}roceedings of the
\uppercase{KIAS-APCTP} \uppercase{I}nternational \uppercase{S}ymposium in 
\uppercase{A}stro-\uppercase{H}adron \uppercase{P}hysics
\uppercase{C}ompact \uppercase{S}tars: \uppercase{Q}uest for 
\uppercase{N}ew \uppercase{S}tates of \uppercase{D}ense \uppercase{M}atter,
\uppercase{N}ov. 10-14, 2003, \uppercase{S}eoul, \uppercase{K}orea.
}}

\author{D.~N.~Aguilera,
\footnote{\uppercase{P}artially supported by 
 \uppercase{L}andesgraduiertenf\"orderung von  \uppercase{M}
\uppercase{V}
\uppercase{G}ermany and by the \uppercase{V}irtual \uppercase{I}nstitute of the \uppercase{H}elmholtz \uppercase{A}ssociation ``\uppercase{D}ense \uppercase{H}adronic \uppercase{M}atter and \uppercase{QCD} \uppercase{P}hase \uppercase{T}ransition'' under grant \uppercase{N}o. \uppercase{VH}-\uppercase{V}1-041.}}

\address{Fachbereich Physik, Universit\"at Rostock,
        Universit\"atsplatz 1, 18051 Rostock, Germany\\
        Instituto de F\'{\i}sica Rosario, Bv. 27 de febrero 210 bis,
        2000 Rosario, Argentina\\
         E-mail: deborah@darss.mpg.uni-rostock.de}

\author{ D.~Blaschke}

\address{Fachbereich Physik, Universit\"at Rostock,
        Universit\"atsplatz 1, 18051 Rostock, Germany\\
        Bogoliubov Laboratory for Theoretical Physics, JINR,
        141980 Dubna, Russia\\
        email: david@thsun1.jinr.ru}

\author{ H.~Grigorian}
\address{Fachbereich Physik, Universit\"at Rostock,
        Universit\"atsplatz 1, 18051 Rostock, Germany\\
        Department of Physics, Yerevan State University, Alex
        Manoogian Str. 1, 375025 Yerevan, Armenia\\
        email: hovik@darss.mpg.uni-rostock.de}

\maketitle

\abstracts{ 
An equation of state for 2-flavor quark matter 
(QM) with diquark condensation under the conditions for compact stars 
-$\beta$-equilibrium, charge and color neutrality- is presented. 
Trapped antineutrinos prevent the formation 
of the diquark condensate
at moderate densities above a critical value of the antineutrino chemical potential $\mu_{\bar \nu_e}^c$. The following consequences are presented:
1) The star develops a 2-phase structure ($\mu_{\bar \nu_e}\geq 
\mu_{\bar \nu_e}^c$):
a color superconducting QM core and a normal QM shell.
2)During the cooling, when the temperature is small enough  ($T<1$ MeV)
the antineutrino mean free path 
becomes larger than the thickness of the normal QM shell and the 
antineutrinos get untrapped in a sudden burst.
The energy release is estimated as $\simeq 10^{52}$ erg and an 
antineutrino pulse is expected to be observed.
}

\section{Introduction}
It has been proposed that cold dense quark matter 
should be in a superconducting state with the formation of a diquark condensate\cite{Alford:2000sx,Blaschke:uj}.
The consequences of the diquark condensation for the 
configuration and the cooling behaviour of compact stars  
have been broadly studied \cite{Blaschke:1999qx,Page:2000wt,Blaschke:2000dy,Blaschke:2003yn} and the question
if this phase can be detected by the signatures 
still remains \cite{Blaschke:2003rg}.

Also the engine for the most energetic phenomena in the universe
 like supernova explosions and gamma ray burst does not have a satisfactory 
explanation yet \cite{Piran:2002gc}and it has been proposed  
that the energy involved could be related with the 
occurence of the color superdonductivity phase \cite{Ouyed:2001cg,Hong:2001gt}. 

Since the pairing energy gap in quark matter is
of the order of the Fermi energy, the diquark condensation gives a 
considerable contribution to the equation of state (EoS) that is estimated
 of the order of $({\Delta}/{\mu})^2$. Disregarding relativistic effects, 
the total binding energy release in the core of a cooling protoneutron star
has been estimated as $({\Delta}/{\mu})^2M_{{\rm core}}\simeq~10^{52}$ erg. 
But, if relativistic effects are considered, the gravitational mass defect 
of the cooling star decreases  when diquark condensation is included 
and there is no explosive process \cite{Blaschke:2003yn} possible since 
the color superconductivity transition is second order.

In this work a new mechanism of releasing the energy in an explosive way is presented (for the original idea see \cite{Aguilera:2002dh}). During the collapse of a protoneutron star antineutrinos are produced by the $\beta$-processes and remain trapped due to the small mean free path.
This increases the asymmetry in the system and therefore the diquark condensation is inhibited at moderate densities. 
So, a two-phase structure developes in the star: 
a superconducting interior and a sorrounding shell of normal quark matter, 
the latter being opaque to  antineutrinos for $T\geq 1$ MeV \cite{Reddy:1997yr}. 
In the cooling process the antineutrino mean free path
increases above the size of this normal
matter shell and  an outburst of neutrinos
occurs releasing an energy of about $10^{51}-10^{52}$ erg. This 
first order phase transition leads to an explosive phenomenon in which 
a pulse of antineutrinos could be observed.

\subsection{Equation of state for 2-flavour quark matter}

A nonlocal chiral quark model for 2-flavour $\{u,d\}$ and three color 
$\{r,b,g\}$
superconducting (2SC) 
quark matter in the mean field approximation is used, for details see 
\cite{Blaschke:2003yn,Grigorian:2003vi}. The order parameters are the mass gap $\phi_f$ and the
diquark gap $\Delta$ for the chiral and superconducting phase transitions
 respectively.
As in \cite{Grigorian:2003vi}, the following chemical potentials are introduced: $\mu_q =  (\mu_u+\mu_d)/2$  for  quark number, 
$\mu_I = (\mu_u-\mu_d)/2$ for isospin asymmetry and  $\mu_8$ 
for color charge asymmetry. The deviation in the color space is considered 
$\mu_8 \ll \mu_q$, so the effect of considering $\mu_8$ is neglected.

The quark thermodynamic potential is expresed as \cite{Kiriyama:2001ud}
\begin{eqnarray}
\lefteqn{\Omega_q(\phi,\Delta;\mu_q,{\mu_I},T)+\Omega_{vac} = 
\frac{\phi^2}{4G_1}+
\frac{\Delta^2}{4G_2}}
\nonumber\\
& & 
-
\frac{2}{2\pi^2}\int^\infty_0dqq^2(N_c-2)\{2E_{\phi} 
+\omega~[E_{\phi}-\mu_q-{\mu_I},T] 
\nonumber\\
& & 
+\omega~[E_{\phi}-\mu_q+{\mu_I},T]
+\omega~[E_{\phi}+\mu_q-{\mu_I},T]
+\omega~[E_{\phi}+\mu_q+{\mu_I},T]
\}\nonumber\\
& &
-\frac{4}{2\pi^2}\int^\infty_0dqq^2\{E_{+}+E_{-}
+\omega[E_{\phi}^{-}-{\mu_I},T]
\nonumber\\
& &
+\omega[E_{\phi}^{-}+{\mu_I},T]
+\omega[E_{\phi}^{+}-{\mu_I},T]
+\omega[E_{\phi}^{+}+{\mu_I},T]
\}
\end{eqnarray}
with
\begin{eqnarray}
\omega[E,T] = T\ln\left[1+\exp\left(-E/T\right)\right]~.
\end{eqnarray}

The dispersion relations for the quarks of unpaired  and paired colors are respectively,
\begin{eqnarray}
{E_\phi}^2&=& q^2+{(m+F^2(q)\phi)}^2\\ 
{E_\phi^\pm}^2&=&(E_{\phi}\pm\mu)^2+F^4(\bar q){\Delta}^2 
\end{eqnarray}

The interaction between the quarks
is implemented via a Gaussian formfactor function 
$F(q)$ in the momentum space (Gaussian types give stable hybrid configurations 
\cite{Blaschke:2003rg}) as  $F(q)=\exp(-q^2/\Lambda^2)~$.

The parameters $\Lambda = 1.025$ GeV, $G_1= 3.761~\Lambda^2$  and
$m_u=m_d=m=2.41$ MeV are fixed by the pion mass,
pion decay constant and
the constituent quark mass at $T=\mu=0$ \cite{Schmidt:1994di}. 
The constant $G_2$ is a free parameter of the approach and is fixed as 
$G_2=0.86~ G_1$.

\subsubsection{Stellar matter conditions}

The stellar matter in the quark core  of compact stars is considered 
to consists of 
$u$ and  $d$ quarks and leptons  
(electrons $e^-$  and antineutrinos $\bar \nu_e$)
 under the following conditions

\begin{itemize}
\item
$\beta$-equilibrium $\quad\quad d \longleftrightarrow u+e^-+\bar \nu_e,$ 
$\quad\quad \mu_e +\mu_{\bar \nu_e} = -2\mu_I,$
\item
Charge neutrality $\quad\quad \frac{2}{3}n_u-\frac{1}{3}n_d-n_e = 0,$
$\quad\quad
n_B+n_I-2n_e = 0,$
\item
Color neutrality $\quad\quad n_8=0,$
$\quad\quad
2n_{qr}-n_{qb}=0,$
\end{itemize}
where 
$n_j= \frac{\partial \Omega}{\partial \mu_j}\bigg|_{T,\phi=\phi_0,\Delta=\Delta_0}$ are the number densities corresponding to the chemical potential 
$\mu_j$ defined above. 

The lepton contributions ($l=e,\bar \nu_e$)
as ideal Fermi gases 
\begin{eqnarray}
\Omega_l(\mu,T)= -\frac{1}{12\pi^2}\mu^4 -\frac{1}{6}\mu^2T^2 
-\frac{7}{180}\pi^2T^4
\end{eqnarray}
are added to the quark thermodynamical potential
\begin{eqnarray}
\Omega(\phi,\Delta;\mu_q,\mu_I,\mu_e,T) = \Omega_q(\phi,\Delta;\mu_q,\mu_I,T)+\Omega_e(\mu_e,T)+\Omega_{\bar \nu_e}(\mu_{\bar \nu_e},T).
\end{eqnarray}

The baryon chemical potential $\mu_B = 3\mu_q-\mu_I$ is introduced as the conjugate of the baryon number density  $n_B$.

\begin{figure}[h]
\centerline{
\psfig{figure=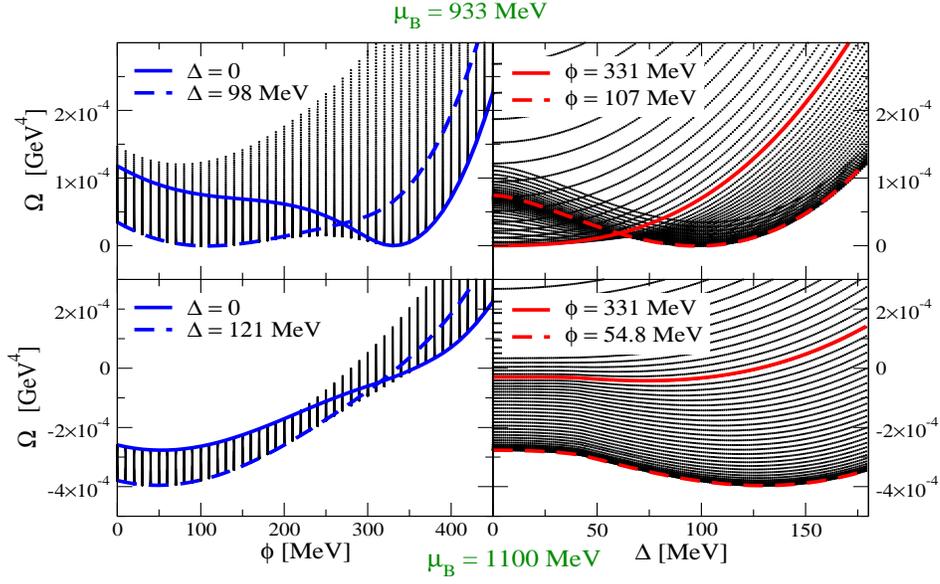,height=9cm,width=13cm,angle=-90}}
\caption{Cuts of the thermodynamic potential $\Omega(\phi,\Delta;\mu_B,\mu_I,T=0)$ in the planes $\Delta = const$ (on the left) and $\phi = const$ (on the right) for two different constant values of $\mu_B$ and the corresponding $\mu_I$. 
For $\mu_B = 933$ MeV (upper panel) two degenerate minima can coexist 
at the values:
$\phi=331$ MeV, $\Delta=0$ (solid lines) and $\phi=107$ MeV, $\Delta=98$ MeV (dashed lines). For $\mu_B = 1100$ MeV (lower panel) the minimum with a nonvanishing diquark $\Delta=121$ MeV and $\phi=54.8$ MeV (dashed lines) is preferable. 
This corresponds to a first order transition from the vacuum to a superconducting phase. In this example $G_2/G_1=1$ was taken.}
\label{Ome_cuts}
\end{figure}

The $\Omega$  function can have several minima in the 
$\phi$, $\Delta$ plane, an example is shown is Fig. \ref{Ome_cuts}.
The global minimum represents the  stable equilibrium of the system and the 
minima search is perfomed solving the gap equations

\begin{eqnarray}
{\partial \Omega \over \partial \phi}\bigg|_{\phi=\phi_0;\Delta=\Delta_0}={\partial \Omega \over \partial \Delta}\bigg|_{\phi=\phi_0;\Delta=\Delta_0}=0
\end{eqnarray}

under the conditions that are mentioned above for the stellar interior.

The thermodynamics of the system, e.g. pressure $P$, energy density $\epsilon$, number density $n$ and entropy density
$s$, is defined via this global minimum

\begin{eqnarray}
\Omega(\phi_{0},\Delta_{0};\mu_B,\mu_I, T)
= \epsilon -Ts -\mu_B n_B -\mu_I n_I  = -P~.
\end{eqnarray}

\begin{figure}[h]
\vspace{-0.5cm}
\centerline{
\psfig{figure=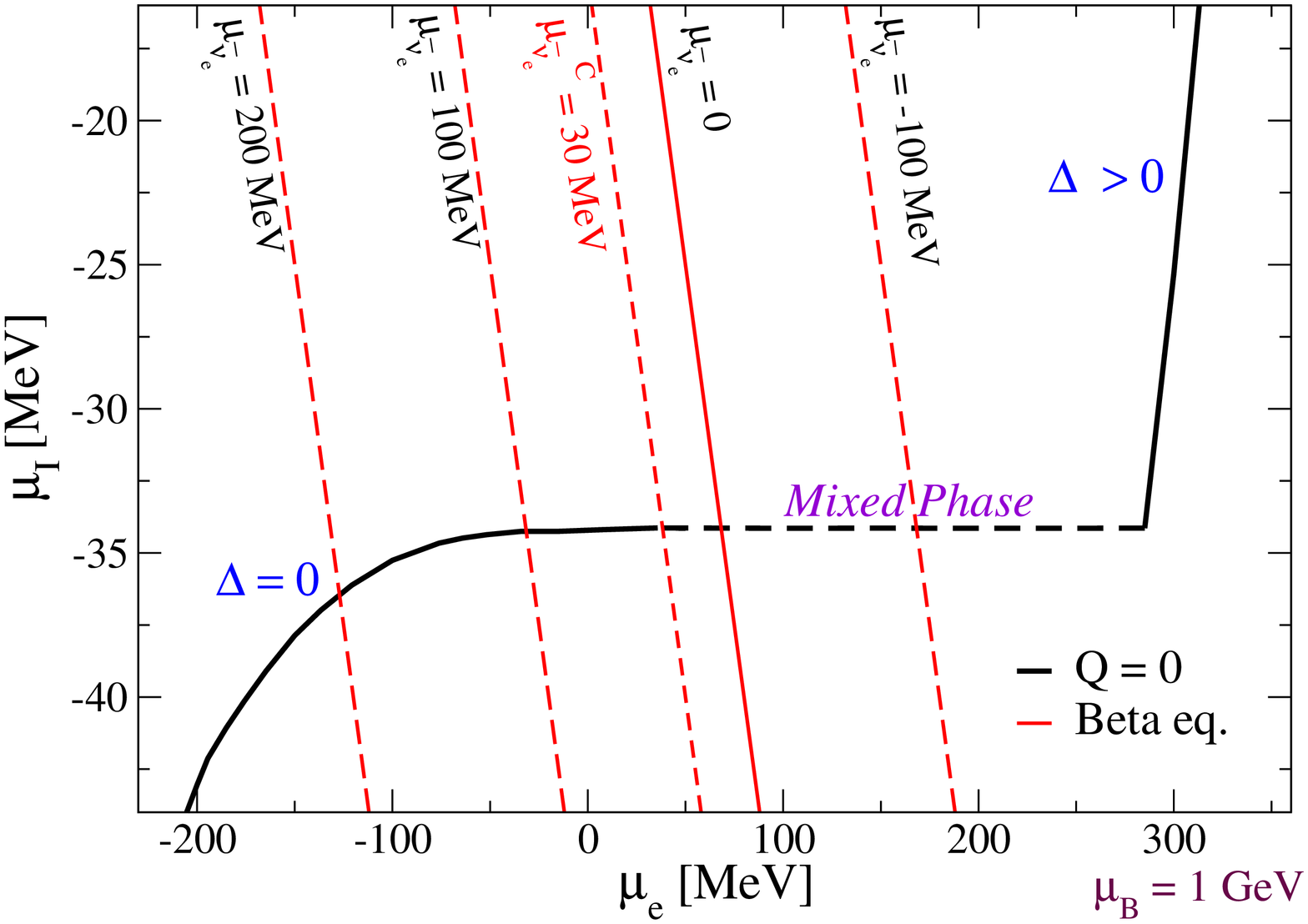,height=7cm,width=7cm,angle=-90}
\psfig{figure=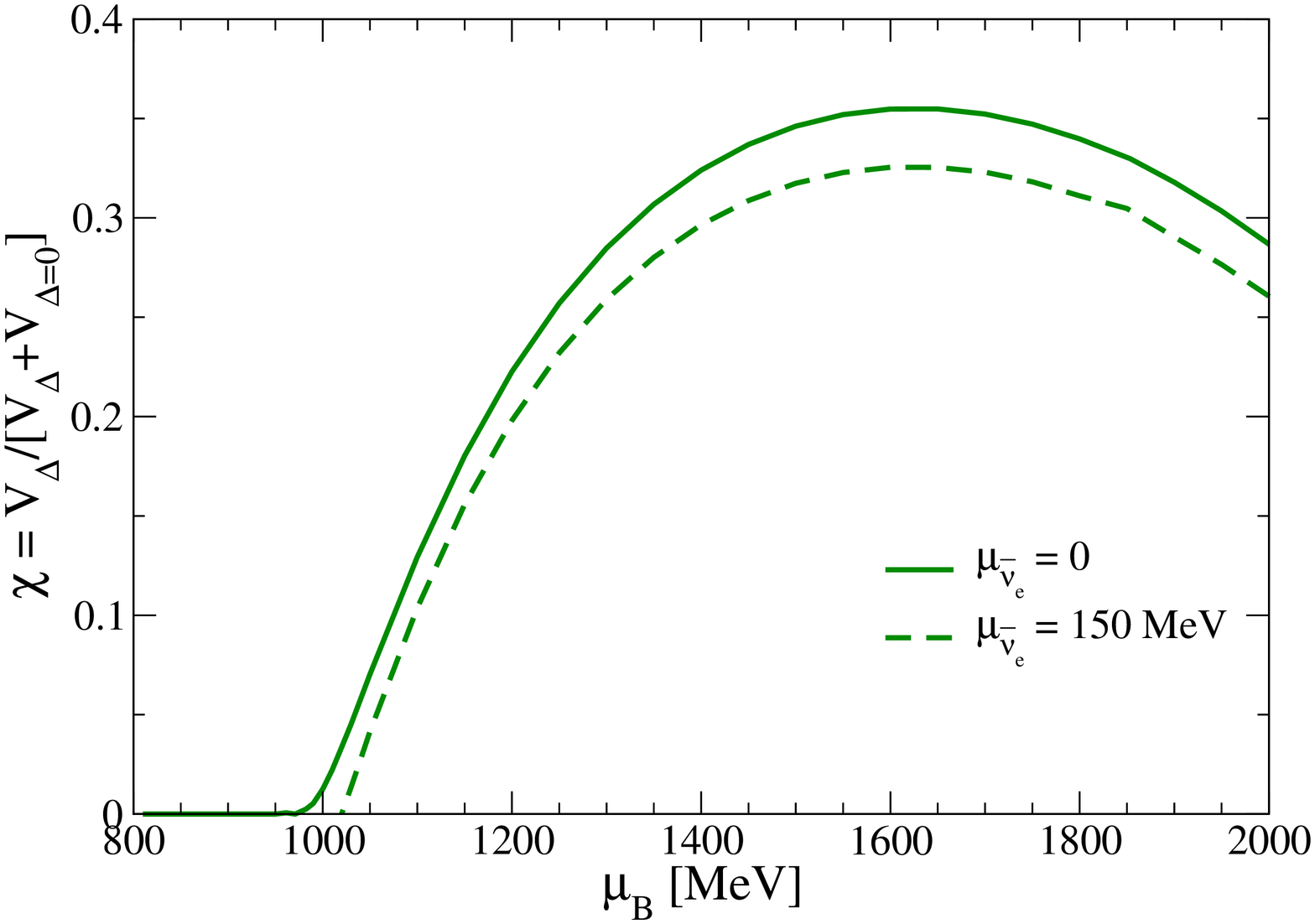,height=7cm,width=7cm,angle=-90}}
\caption{{\bf Left}: Solutions of the gap equations and the charge neutrality condition (solid black line) in the $\mu_I$ vs. 
$\mu_e$ plane. Two branches are shown: states with diquark condensation on the upper right ($\Delta>0$) and states from normal quark matter ($\Delta=0$) on the lower left. The plateau in between corresponds to a mixed phase. 
The lines for the $\beta$-equilibrium condition are also shown (solid and dashed red lines) for different values of the (anti)-neutrino chemical potential. The stellar matter should satisfy both conditions (intersection of the corresponding lines) and therefore for $\mu_{\bar \nu_e}=0$ a mixed phase is preferable.
{\bf Right}:Volume fraction $\chi$ of the phase with nonvanishing
     diquark condensate obtained by a Glendenning construction of
a charge-neutral mixed phase. Results are shown for two different 
values of $\mu_{\bar \nu_e}$.}
\label{fig:volfraction}
\end{figure}

To fulfill the charge neutrality condition (see Fig. \ref{fig:volfraction}, right) a mixed phase  between a subphase with diquark condensation (subscript $\Delta>0$) and normal quark matter subphase (subscript $\Delta=0$) is defined 
via the Glendenning construction. The Gibbs condition for equilibrium 
at fixed $T$ and $\mu_B$ is that the pressure of the subphases should be the same
 
\begin{eqnarray}
P =P^{\Delta>0}(\mu_B,\mu_I,\mu_e,T)=P^{\Delta=0}(\mu_B,\mu_I,\mu_e,T)~.
\end{eqnarray}

The volume fraction $\chi$ that is occupied by the subphase with diquark condensation
 is defined by the charge $Q$ in the subphases,
\begin{eqnarray}
\chi = Q_{\Delta>0}/(Q_{\Delta>0}-Q_{\Delta=0})~,
\end{eqnarray}

and is plotted on the right panel in Fig. \ref{fig:volfraction} for 
different antineutrino chemical potentials as a function of  $\mu_B$. 
In the same way, the  number densities for the different particle species $j$ 
 and the energy density are given by
\begin{eqnarray}
n_j = \chi n_{j_{\Delta>0}}+ (1-\chi)n_{j_{\Delta=0}}~,\\
\epsilon = \chi \epsilon_{\Delta>0}+ (1-\chi)\epsilon_{\Delta=0}~.
\end{eqnarray}
 
\subsubsection{Equation of state with trapped antineutrinos}

Increasing the antineutrino chemical potential $\mu_{\bar \nu_e}$ increases the asymmetry in the system and this shifts the onset of the superconducting phase transition to higher densities.

\begin{figure}[h]
\vspace{-0.5cm}
\centerline{
\psfig{figure=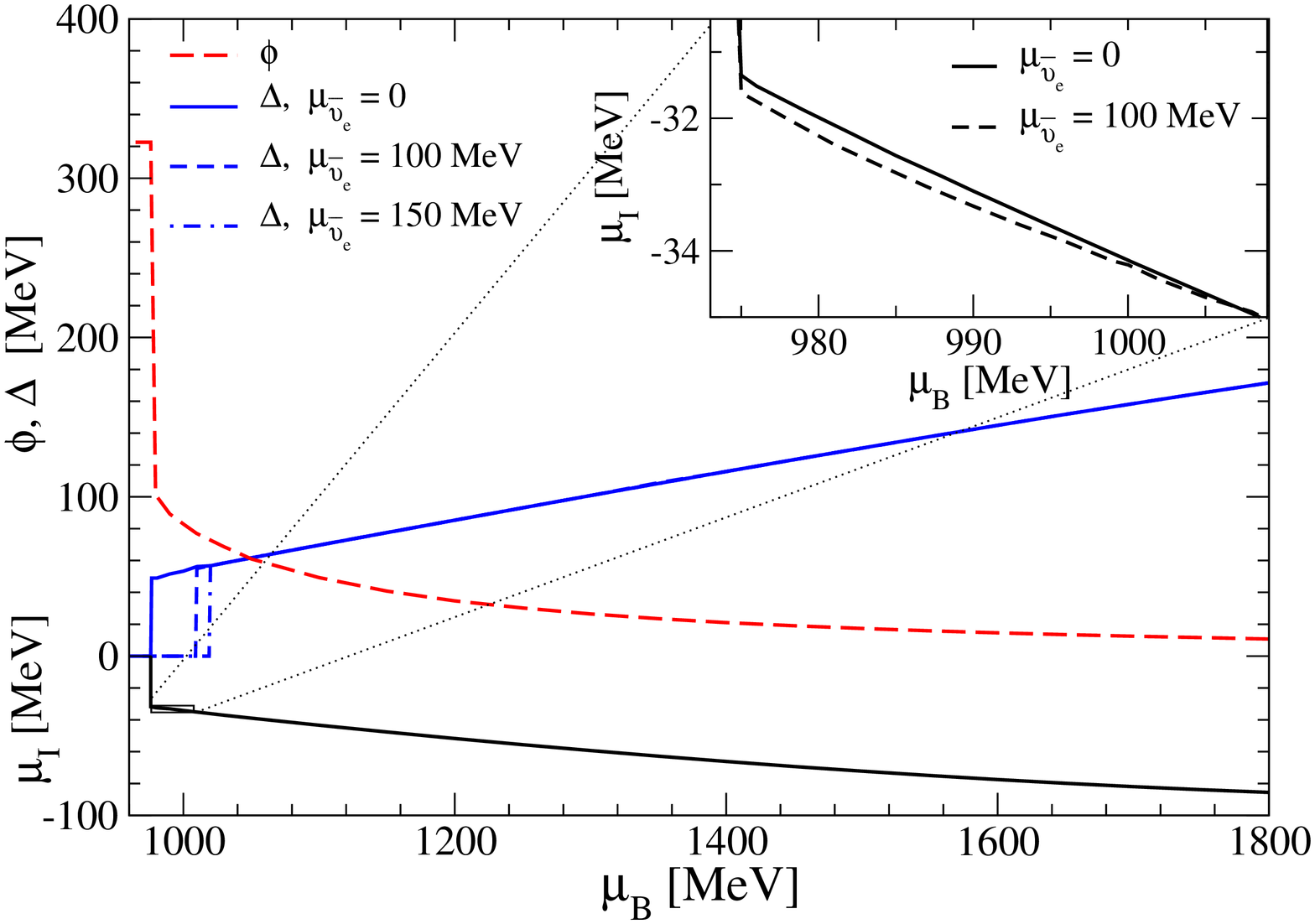,height=7cm,width=7cm,angle=-90}
\psfig{figure=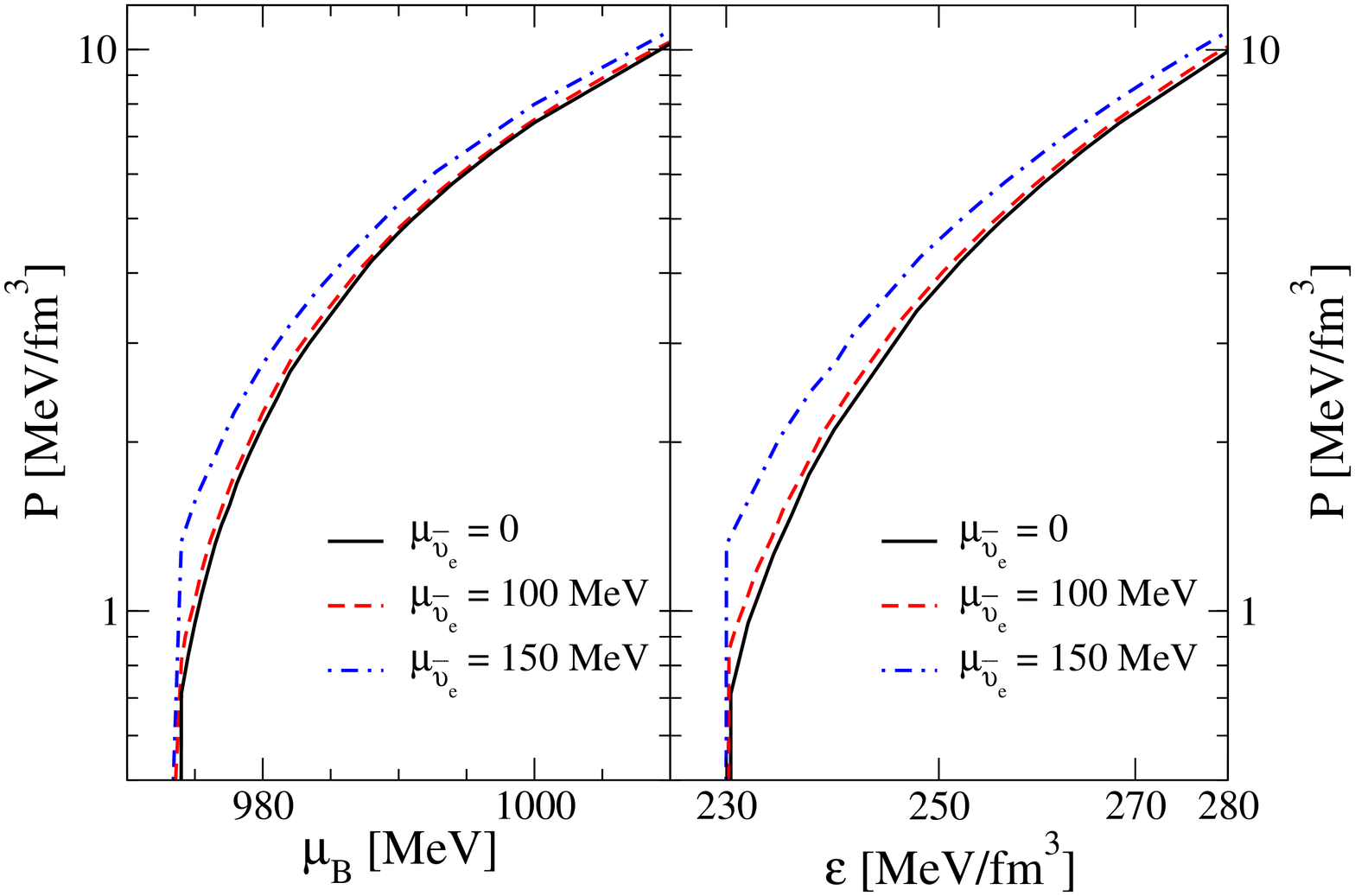,height=7cm,width=7cm,angle=-90}}
\caption{{\bf Left}: Solutions of the gap equations and $\mu_I$  as a function of $\mu_B$. Increasing the antineutrino chemical potential increases the asymmetry in the system and the superconducting phase is inhibited at moderates densities.
{\bf Right}: Equation of state for different values of $\mu_{\bar \nu_e}$ of 
trapped antineutrinos. As $\mu_{\bar \nu_e}$ increases the equation of state becomes harder. 
}
\label{fig:GE}
\end{figure}

Above a critical value of 
$\mu_{\bar \nu_e}\geq \mu_{\bar \nu_e}^c \simeq 30$ MeV (see critical value in Fig. \ref{fig:volfraction}, on the left)  first 
a normal quark matter phase occurs and  then the phase transition to superconducting matter takes place, see Fig. \ref{fig:GE}, on the left.

The consequences for the equation of state can be seen on the right of the Fig. \ref{fig:GE}: the onset of the superconductivity in quark matter
 is shifted to higher densities and the equation of state becames harder. 
  
\subsection{Quark stars and antineutrino trapping}

The configurations for the quark stars are obtained by solving the Tolman-Oppenheimer-Volkoff equations for a set of central quark number densities $n_q$ for which the stars are stable. 

\begin{figure}[h]
\vspace{-0.5cm}
\centerline{
\psfig{figure=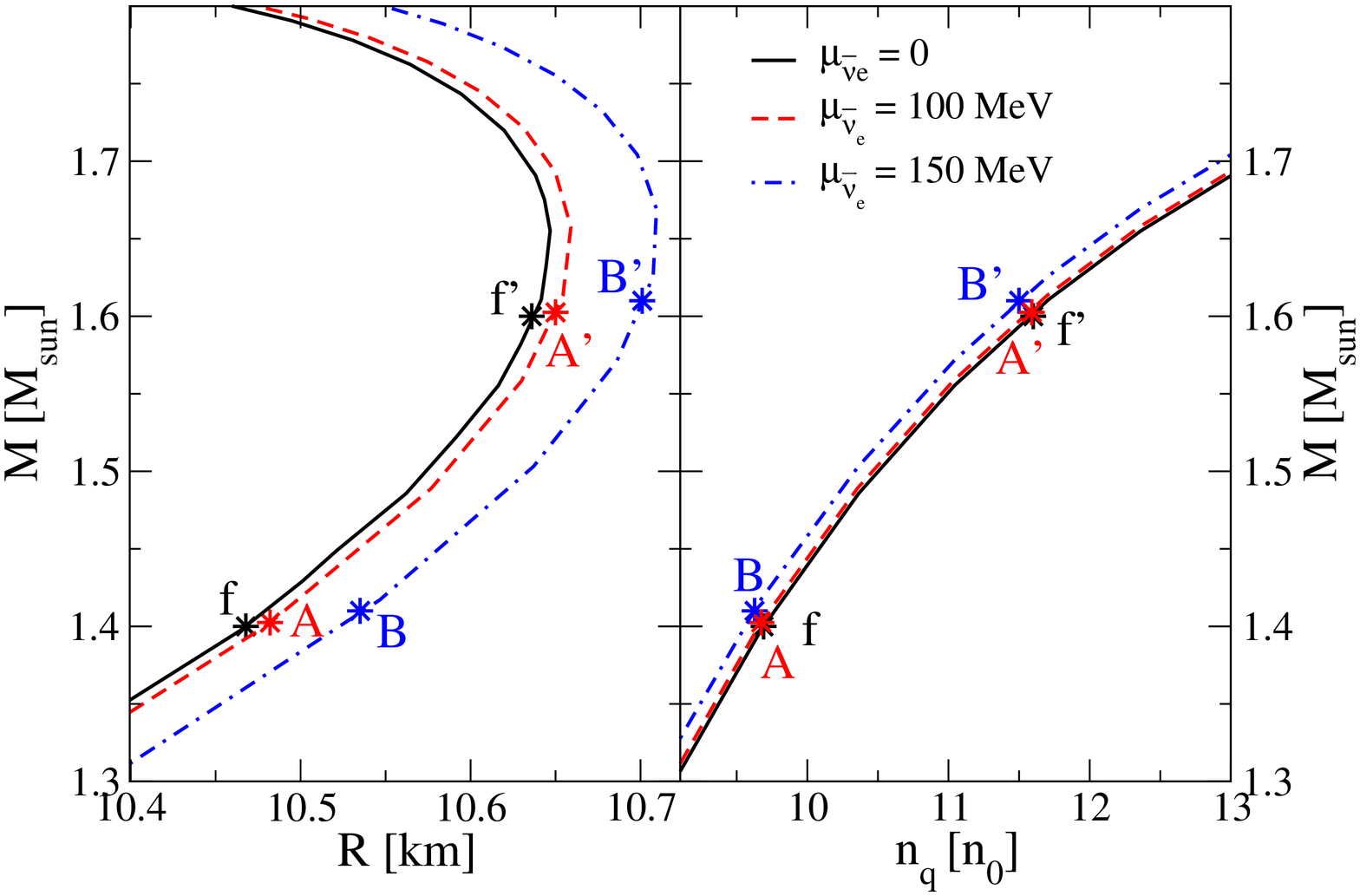,height=7cm,width=7cm,angle=-90}
\psfig{figure=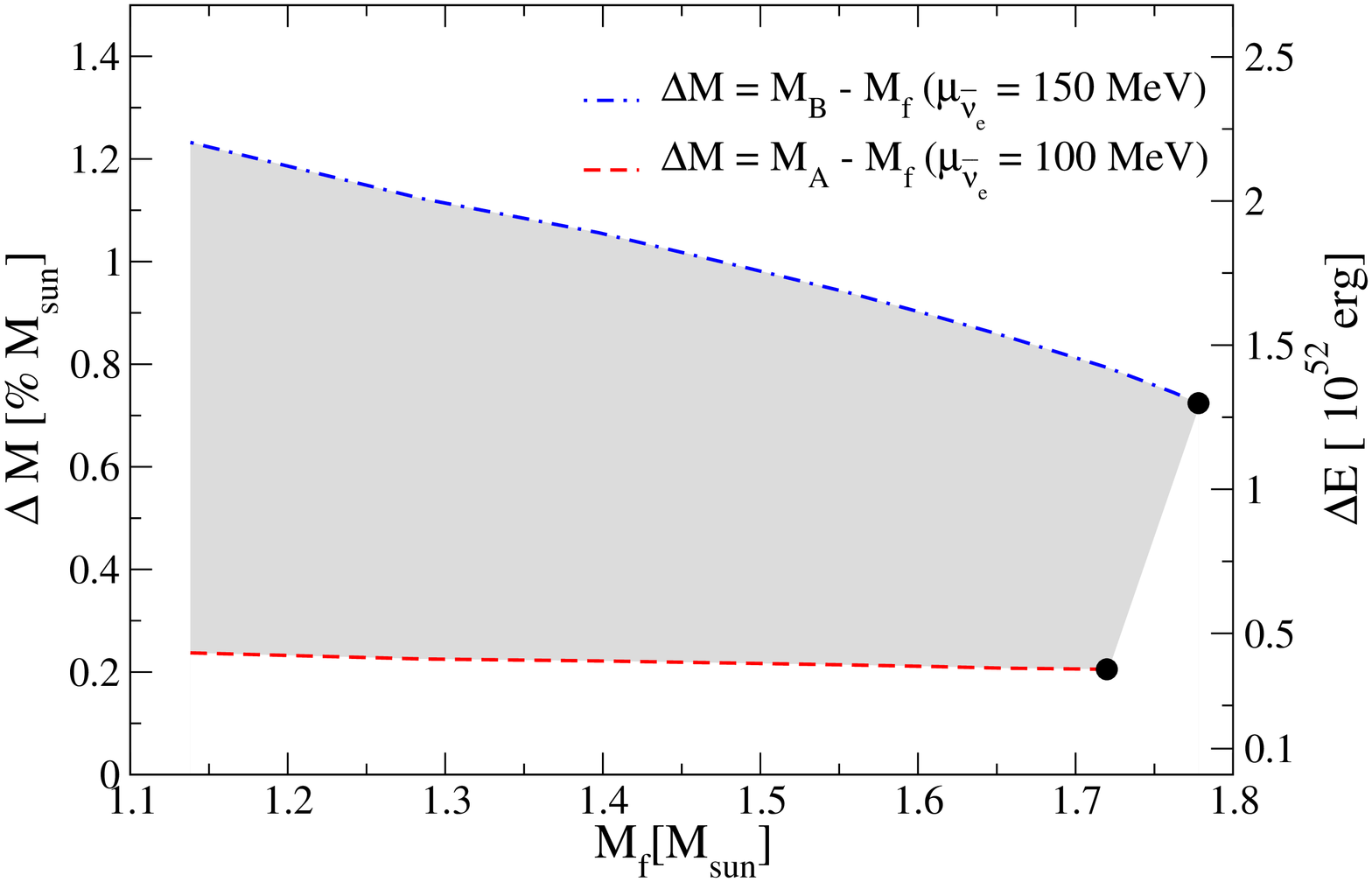,height=7cm,width=7cm,angle=-90}}
\caption{{\bf Left}: Quark star configurations for different antineutrino
chemical potentials $\mu_{\bar \nu_e}=0, ~100,~150$ MeV.
The total mass $M$ in solar masses
($M_{{\rm sun}}\equiv M_\odot$ in the text)
is shown as a function of the radius $R$ (left panel) and
of the central number density  $n_q$ in units of the
nuclear saturation density $n_0$ (right panel).
Asterisks denote two different sets of configurations (A,B,f) and (A',B',f')
with a fixed total baryon number of the set. 
{\bf Right}: Mass defect $\Delta M$ and corresponding energy release $\Delta E$
due to antineutrino untrapping
as a function of the mass of the final state $M_f$.
The shaded region is defined by the estimates for the upper and lower
limits of the antineutrino chemical potential
in the initial state $\mu_{\bar \nu_e}=150$ MeV (dashed-dotted line) and
$\mu_{\bar \nu_e}=100$ MeV (dashed line), respectively.
}
\label{fig:Conf}
\end{figure}

In Fig. \ref{fig:Conf} the configurations for different antineutrino chemical potentials are  shown. The equations of state with trapped antineutrinos are softer and therefore this  allows more compact configurations. The presence of antineutrinos tends to increase the mass of the star for a given central density. 

A reference configuration with total baryon number  $N_B = 1.51~ N_{\odot}$ (where $N_{\odot}$ is the total baryon number of the sun) is chosen and the case with (configurations $A$ and $B$ in Fig. \ref{fig:Conf}) and without antineutrinos ($f$ in  Fig. \ref{fig:Conf}) are compared. 
A mass defect can be calculated between the configurations with  and without 
trapped antineutrinos at constant total baryon number and the result is shown on the right panel of  Fig. \ref{fig:Conf}). The mass defect could be interpreted as an energy release if the configurations $A,B$ with antineutrinos are initial states and the configuration $f$ without them is the final state of a protoneutron star evolution. 

\subsubsection{Protoneutron star evolution with antineutrino trapping}

\begin{figure}[h]
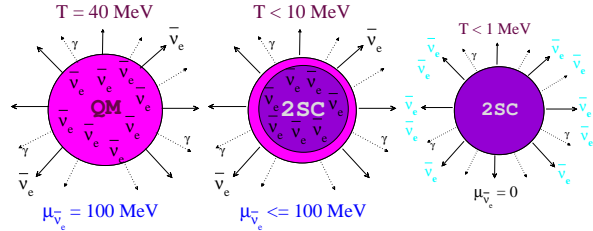

\vspace{-0.5cm}
\centerline{
\parbox{6cm}{\psfig{figure=esf1.epsi,width=2.5cm,angle=-90}}
\parbox{6cm}{\psfig{figure=esf2.epsi,width=2.5cm,angle=-90}}
\parbox{6cm}{\psfig{figure=esf3.epsi,width=2.5cm,angle=-90}}}
\caption{Left graph: Quark star cooling by antineutrino and photon emission from the surface. Middle graph: Two-phase structure developes due to the trapped antineutrinos: a normal quark matter shell and a superconducting interior. Right graph: Antineutrino untrapping and burst-type release of energy.}
\label{fig:kugel}
\end{figure}

\begin{figure}[h]
\vspace{-0.5cm}
\centerline{
\psfig{figure=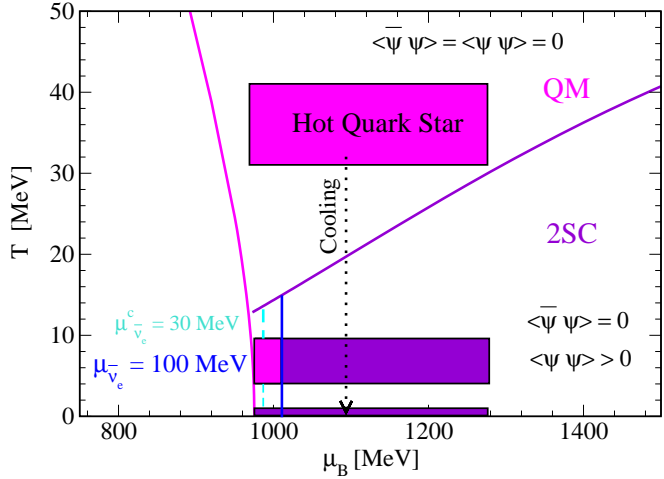,width=10cm,angle=-90}
}
\caption{Star evolution corresponding to Fig. \ref{fig:kugel} plotted in the phase diagram}
\label{fig:phasediag}
\end{figure}

After the collapse of a protoneutron star the star cools down 
by surface emission of photons and antineutrinos. Antineutrinos are trapped 
because they were generated by the direct $\beta$-process in the hot and dense 
matter and could not escape due to their small mean free path.
The region of the star where the temperature falls below the density dependent
critical value for diquark condensation, will transform to the color
superconducting state which is almost transparent to (anti)neutrinos.
But nevertheless due to the trapped antineutrinos there is a dilute normal
quark matter shell which prevents neutrino escape from the
superconducting bulk of the star, see Fig. \ref{fig:kugel} and Fig. 
\ref{fig:phasediag}.
The criterion for the  neutrino untrapping transition is to cool the star
below a temperature of about 1 MeV when the mean free path of neutrinos
becomes larger than the shell radius \cite{Prakash:2001rx}.
If at this temperature the antineutrino chemical potential is still large
then the neutrinos can escape in a sudden outburst. If it is small then there
will be only a gradual increase in the luminosity.
An estimate for the possible release of energy within the outburst scenario
can be given via the mass defect  defined in the previous subsection
between an initial configuration with trapped neutrinos (state $A$ or $B$) and
a final configuration without neutrinos (state $f$).

\section{Conclusions}

The effects of trapped antineutrinos on the diquark condensates in quark star configurations are investigated. At fixed baryon number the energy release in the antineutrino untrapping transition is of the order of $10^{52}$ erg. 
This is a first order transition and leads to an explosive release of energy
 that could help to explain energetic phenomena in the universe like gamma ray bursts or supernova explosions.

\section*{Acknowledgments}
D.N.A and D.B. thank the organizers of the KIAS-APCTP International 
Symposium in Astro-Hadron Physics of 
Compact Stars
for their invitation and in particular 
D.K. Hong and the staff of the Department of Physics 
at Pusan National Univertity for their hospitality and interest in this work. 
D.N.A acknowledges the DAAD-HOST programm D/03/31497 and 
the local Department of Physics 
for the financial support of the
visit to Pusan University.
The authors enjoyed the wonderful atmosphere and lively discussions with 
all colleagues
during the conference.



\begin{thebibliography}{0}

\bibitem{Alford:2000sx}
M.~Alford, J.~A.~Bowers and K.~Rajagopal,
J.\ Phys.\ G {\bf 27} (2001) 541
[Lect.\ Notes Phys.\ {\bf 578} (2001) 235]
[arXiv:hep-ph/0009357].

\bibitem{Blaschke:uj}
D.~Blaschke, N.~K.~Glendenning and A.~Sedrakian,
``Physics Of Neutron Star Interiors.'' Springer Lecture Notes in Physics {\bf 578} (2001).  
{\it Prepared for ECT* International Workshop on Physics of Neutron Star Interiors (NSI00), Trento, Italy, 19 Jun - 7 Jul 2000}

\bibitem{Blaschke:1999qx}
D.~Blaschke, T.~Klahn and D.~N.~Voskresensky,
Astrophys.\ J.\  {\bf 533} (2000) 406
[arXiv:astro-ph/9908334].

\bibitem{Page:2000wt}
D.~Page, M.~Prakash, J.~M.~Lattimer and A.~Steiner,
Phys.\ Rev.\ Lett.\  {\bf 85} (2000) 2048
[arXiv:hep-ph/0005094].

\bibitem{Blaschke:2000dy}
D.~Blaschke, H.~Grigorian and D.~N.~Voskresensky,
Astron.\ Astrophys.\  {\bf 368} (2001) 561
[arXiv:astro-ph/0009120].

\bibitem{Blaschke:2003yn}
D.~Blaschke, S.~Fredriksson, H.~Grigorian and A.~M.~Oztas,
[arXiv:nucl-th/0301002].

\bibitem{Blaschke:2003rg}
D.~Blaschke, H.~Grigorian, D.~N.~Aguilera, S.~Yasui and H.~Toki,
AIP Conf.\ Proc.\ {\bf 660} (2003) 209
[arXiv:hep-ph/0301087].


\bibitem{Piran:2002gc}
T.~Piran and E.~Nakar,
arXiv:astro-ph/0202403.

\bibitem{Ouyed:2001cg}
R.~Ouyed and F.~Sannino,
Astron.\ Astrophys.\ {\bf 387} (2000) 725
[arXiv:astro-ph/0103022].



\bibitem{Hong:2001gt}
D.~K.~Hong, S.~D.~H.~Hsu and F.~Sannino,
Phys.\ Lett.\ B {\bf 516}, (2001) 362
[arXiv:hep-ph/0107017].


\bibitem{Aguilera:2002dh}
D.~Aguilera,  D.~Blaschke and H.~Grigorian,
[arXiv:astro-ph/0212237].


\bibitem{Reddy:1997yr}
S.~Reddy, M.~Prakash and J.~M.~Lattimer,
Phys.\ Rev.\ D {\bf 58} (1998) 013009
[arXiv:astro-ph/9710115].

\bibitem{Grigorian:2003vi}
H.~Grigorian, D.~Blaschke and D.~N.~Aguilera,
arXiv:astro-ph/0303518.

\bibitem{Kiriyama:2001ud}
O.~Kiriyama, S.~Yasui and H.~Toki,
Int.\ J.\ Mod.\ Phys.\ E {\bf 10} (2001) 501
[arXiv:hep-ph/0105170].

\bibitem{Schmidt:1994di}
S.~M.~Schmidt, D.~Blaschke and Y.~L.~Kalinovsky,
Phys.\ Rev.\ C {\bf 50} (1994) 435.

\bibitem{Prakash:2001rx}
M.~Prakash, J.~M.~Lattimer, R.~F.~Sawyer and R.~R.~Volkas,
Ann.\ Rev.\ Nucl.\ Part.\ Sci.\  {\bf 51} (2001) 295
[arXiv:astro-ph/0103095].

\end{thebibliography}
\end{document}